\documentclass[a4paper]{article}

\usepackage[english]{babel}
\usepackage[latin1]{inputenc}
\usepackage[T1]{fontenc}

\usepackage[a4paper,top=3cm,bottom=2cm,left=3cm,right=3cm,marginparwidth=1.75cm]{geometry}

\usepackage{amsmath}
\usepackage{graphicx}
\usepackage[colorinlistoftodos]{todonotes}
\usepackage[colorlinks=true, allcolors=blue]{hyperref}

\newcommand{\inst}[1]{\textsuperscript{#1}}

\begin{document}

\title{\Huge Dynamics of new strain emergence\\
on a temporal network}{}
\date{}
\maketitle

\begin{center}
Sukankana Chakraborty\inst{1}, 
Xavier R. Hoffmann\inst{2,3},
Marc G. Leguia\inst{4,5},
Felix Nolet\inst{6},
Elisenda Ortiz\inst{2,3},
Ottavia Prunas\inst{7,8,9},
Leonardo Zavojanni\inst{10},
Eugenio Valdano\inst{11},
Chiara Poletto\inst{12}

\vspace{1cm}
{\footnotesize
\inst{1}Department of Electronics and Computer Science (ECS), University of Southampton, Highfield Campus, SO17 1BJ, Southampton, United Kingdom.
\inst{2}Departament de F\'isica de la Mat\`eria Condensada, Universitat de Barcelona, Spain. 
\inst{3}Universitat de Barcelona Institute of Complex Systems (UBICS), Spain.
\inst{4}Faculty of Information Studies, Novo Mesto, Slovenia.
\inst{5}Department of Information and Communication Technologies, Universitat Pompeu Fabra, Barcelona, Catalonia, Spain. 
\inst{6}Laboratory of Dynamics in Biological Systems; Department of Cellular and Molecular Medicine, KU Leuven; Leuven, Belgium. 
\inst{7}ISI Foundation; Via Chisola 5; 10126; Torino; Italy.
\inst{8}Complex Systems for Life Sciences, Universit\'a degli Studi di Torino, 10124 Torino, Italy.
\inst{9}GSK Vaccines, 53100 Siena, Italy.
\inst{10}Section for the Science of Complex Systems; CeMSIIS; Medical University of Vienna; Spitalgasse 23; A-1090; Vienna, Austria.
\inst{11}Departament d'Enginyeria Informatica i Matematiques, Universitat Rovira i Virgili.
\inst{12}Sorbonne Universit\'es, UPMC Univ. Paris 06, INSERM, Institut Pierre Louis d'Epid\'emiologie et de Sant\'e Publique (IPLESP UMR-S 1136), 75012 Paris, France.
}
\end{center}

\begin{abstract}
Multi-strain competition on networks is observed in many contexts, including infectious disease ecology, information dissemination or behavioral adaptation to epidemics.
Despite a substantial body of research has been developed considering static, time-aggregated networks, it remains a challenge to understand the transmission of concurrent strains when links of the network are created and destroyed over time.
Here we analyze how network dynamics shapes the outcome of the competition between an initially endemic strain and an emerging one, when both strains follow a susceptible-infected-susceptible dynamics, and spread at time scales comparable with the network evolution one.
Using time-resolved data of close-proximity interactions between patients admitted to a hospital and medical health care workers, we analyze the impact of temporal patterns and initial conditions on the dominance diagram and coexistence time. We find that strong variations in activity volume cause the probability that the emerging strain replaces the endemic one to be highly sensitive to the time of emergence. The temporal structure of the network shapes the dominance diagram, with significant variations in the replacement probability (for a given set of epidemiological parameters) observed from the empirical network and a randomized version of it.
Our work contributes towards the description of  the complex interplay between competing pathogens on temporal networks.
\end{abstract}

\section{Introduction}

The study of infection spread on contact networks has been the focus of epidemiological research for a significantly long time. Recently, increasing attention is being devoted to studying the co-circulation of multiple infections on the same networked population, and the impact of competitive and cooperative interference among them on the epidemic dynamics \cite{Marceau2010,Karrer2011,Prakash2012,Ghanbarnejad2013,Poletto2013,Sanz2014,Dufresne2015}. An important goal is to elucidate dynamical mechanisms underlying the emergence of new pathogen strains and the associated phenomena of vaccine failure and antibiotic resistance~\cite{Fraser1968,Fox2017}. Applications, however, go beyond infectious disease epidemiology, and include the spread of competing (or cooperating) ideas, memes,computer viruses and products. 

Overall the goal of multi-strain models is the understanding of the co-existence outcome in varying  key epidemiological parameters - e.g. transmissibility and duration of infection period. For instance, it was shown that two competing strains rarely coexist~\cite{Karrer2011,Prakash2012}. The most efficiently spreading strain quickly becomes dominant, and leads the other to extinction. Moreover, a key result is that the conditions on disease parameters leading to dominance often depend on the properties of the network substrate \cite{Marceau2010,Karrer2011,Ghanbarnejad2013,Poletto2013,Sanz2014,Dufresne2015}. 

So far, most works have considered static networks and only very few have accounted for the dynamics of contacts \cite{Pare2017,Rodriguez2017}. The spread of single infections on temporal networks, on the other hand, is being extensively studied, the interest in the subject being prompted by the increasing availability of time resolved contact data~\cite{Gonzalez2008,Bajardi2011}. Face-to-face interactions have been measured by RFID technology in school, workplaces and hospitals, among the others ~\cite{Salathe2010,Isella2011,Stehle2011,Mastandrea2015,Obadia2015}. Time resolved data on sexual encounters are also available~\cite{Rocha2010}. These data highlight the dynamical properties of human interactions, such as node turnover, burstiness, heterogeneous activity potential, presence of temporal motifs and correlations, that were shown to profoundly alter in some cases the epidemic spread ~\cite{Karsai2011,Miritello2011,Holme2012,Perra2012,Holme2016,Valdano2015}.

Here we study the competition between an endemic and an emerging strain on an empirical temporal network of face-to-face contacts collected in a hospital. We reconstruct the outcome of the emergence (either replacement or extinction) in varying transmissibility and infection duration of the emerging strain. We focus on the regime in which time scales of spreading and network evolution are comparable and we consider different times of emergence in correspondence of peaks and drops in activity. Comparing the competition outcomes obtained on the empirical network and on randomized time-shuffled models we describe how contact dynamics alters the likelihood and manner in which the new strain emerges.
 
\section{Methods}

\subsection{Contact network}
We use the network of face-to-face proximity interactions among patients and health care workers in a hospital ward, collected by the SocioPatterns group~\cite{Vanhems2013}. The data contain $N=75$ individuals, and their proximity interactions with the time of occurrence. The data collection period spans 4 days. We aggregate contacts over a 30-minute windows. The resulting temporal network is a sequence of $T=192$ time steps, each encoding the topology of contacts at that time.

\subsection{Pathogen dynamics}
We model both the endemic and the emergent strain through Susceptible-Infectious-Susceptible (SIS) compartmental models. The nodes in the network can be either Susceptible, or infected by either strain. The two strains are mutually exclusive so that concurrent infection of the same nodes by both strains is impossible. At every time step, each Infectious infects a Susceptible neighbor with a probability $\beta_0,\beta$ for the endemic and emergent strain, respectively. Infectious nodes also recover back to the Susceptible state, spontaneously, with probability $\mu_0$ (endemic strain), $\mu$ (emergent strain).

In order to study pathogen emergence and takeover, we set up our stochastic simulations as follows
{\itshape i}) We fix $\beta_0=0.325$ and $\mu_0=0.08$, which corresponds to an average infectious period of $6.5$ hours.
{\itshape ii}) We let the endemic strain evolve to equilibrium, using a relaxation time of 24 hours. {\itshape iii}) We fix the time of emergence $t_{\mathrm{inj}}$ of the emergent strain. We use two values: 7PM on the second day, and 7AM of the third day. We dub these two points AP (after peak) and BP (before peak), respectively. At $t_{inj}$ we seed one Susceptible node with the emergent strain, and let the system evolve until either strain goes extinct. In order to be able to simulate the spread on time scales longer than the available dataset, we impose periodic boundary conditions on the temporal network~\cite{Stehle2011}. 

\subsection{Randomized null model}
In order to gauge the impact of the temporal and topological properties of the contact network, we set up a randomized null model, aimed at breaking the correlations between time and topology. At each time step of the spreading simulation we sample randomly with replacement a network configuration among the $T$ of the original dataset. 

\subsection{Parameter exploration}
We explore values of $\beta$ in the range $[0.1,0.9]$, and $\mu$ in $[0.01,0.125]$, using steps of 0.025, 0.0025, respectively. Then, in order to accumulate sufficient statistics for our measures, we run $10000$ iterations for each parameter configuration.

\section{Results}

By varying the epidemiological parameters of the emergent strain ($\beta,\mu$), we study the phenomenology of pathogen interaction at different parameter values, and investigate the successful replacement of the endemic strain by the emergent one. We also probe the temporal and topological features of the network that drive this outcome.

\subsection{Strain replacement}
For each $(\beta,\mu)$ we compute the {\itshape probability of replacement} ($P_\mathrm{R}$), i.e., the probability that, after emergence, the emergent strain takes over the population and drives the endemic strain to extinction.
Unsurprisingly, Fig.~\ref{figure2} shows that high values of transmissibility $\beta$ favor replacement (high $P_\mathrm{R}$). Replacement, however, is still possible when the transmissibility of the emergent strain is lower than the endemic strain's ($\beta<\beta_0$), provided the emerging leads to low values of recovery probability (long infectious periods). What we find to dramatically impact the behavior of $P_\mathrm{R}$, however, is the time of emergence ($t_{\mathrm{inj}}$). The probability of replacement is systematically higher when the second strain emerges at BP, with respect to AP. Specifically, Fig.~\ref{figure2} shows that emergent strains with short infectious periods (high $\mu$) emerging at AP have a negligible probability of replacement. Conversely, the same strains, injected at BP, get to fixation with much higher probability.
On the other hand, diseases with long infectious periods are less sensitive to the emergence time, as marked by the small $P_\mathrm{R}$ difference in the lower part of Fig.~\ref{figure2}C.
In order to uncover which properties of the temporal network may explain the varying fitness of the emerging strain as the emergence time varies, we perform the spreading simulations on the null model.
Figure~\ref{figure3} depicts the difference in $P_\mathrm{R}$ between the real network and the null model, for emergence at both BP and AP. At AP, the probability of replacement is dramatically higher in the randomized null model than in the real network, with differences of $P_\mathrm{R}$ above $0.2$ for fast-spreading pathogens (high $\beta$). In general, for all of the parameters ($\beta,\mu$) considered, the null model is significantly more prone to replacement than the real network.
However, the picture is substantially different when the strain is injected at BP, and two scenarios are revealed: for long infectious periods (low $\mu$), the null model still shows a higher $P_\mathrm{R}$, while with short ones, replacement ins more likely in the real network.

\subsection{Time of coexistence}
While $P_\mathrm{R}$ tells us how likely the emergent strain is to take over, it gives no information on how it gets to fixation. One simple measure to gauge that is the {\itshape time of coexistence}, i.e., the number of time units in which the prevalence of both strains is higher than zero. Panels {\itshape A,D} in Fig.~\ref{figure4} show that the average time of coexistence is an decreasing function of both $\beta$ and $\mu$, and that holds for both the real network and the null model. This means that, as expected, highly contagious pathogens (large $\beta$) can overtake the network more quickly. 
On the other hand, time of coexistence decreases with $\mu$ since the endemic strain is more likely to persist in regions where the emergent is subcritical, where it is quickly cleared from the system.
In order to further disentangle this, we also compute the average time of coexistence conditioned on the emergent replacing the endemic strain. By comparing panels {\itshape A,B} in Fig.~\ref{figure4} we see that conditioning on the outcome visibly changes the length of coexistence. Overall, the average coexistence time is longer when the emergent wins, as expected as it needs the time to overtake the endemic one, while coexistence with no replacement is often the short transient between emergent strain introduction and spontaneous extinction.
Plots in Fig.~\ref{figure4}({\itshape A,D}) show that randomization does not visibly change the trend of the average time of coexistence, while conditioning on replacement (Fig.~\ref{figure4}({\itshape B,E})) leads to significantly different scenarios in the real network and the null model. The real network shows markedly lower times, and such difference is most visible around the green line. Close to this critical regime, where the emergent strain is not efficient, bottleneck effects induced by the activity pattern of the network select strains that are able to quickly replace the endemic one, inducing a low time of coexistence. In the null model, in the absence of bottlenecks, the critical region witnesses the highest values of time of coexistence, as the disease spreads slowly up to takeover.
In the plots Fig.~\ref{figure4}({\itshape C,F}), for selected values of $\beta,\mu$ close to the critical region, we show the full distribution of the time of coexistence, showing that conditioning on replacement changes not only the average, but the whole shape. A peculiar pattern emerges, where the interplay between the time scale of the disease and of contact activity select specific values of coexistence time. Null model breaks this, and exhibits very similar distribution profiles between conditioning and not conditioning on replacement. This means that, when temporal correlations are broken, the dynamical pathways leading to the dominance of either strain are qualitatively the same.

\section{Discussion \& Conclusion}

We studied the dynamics of emergence of a new strain on a temporal network of contacts in the regime in which the spreading time scales are comparable with the time scale of the network evolution. By comparing results on an empirical network and on a randomized null model we found that the temporal dimension alters the outcome of emergence and shape the dominance diagram. Several studies addressed the role of the network substrate on the competition between strains and network topology and extent of coupling were, in some cases, shown to alter the epidemiological traits enabling one strain to dominate \cite{Prakash2012,Karrer2011,Sanz2014,Dufresne2015,Funk2010,Marceau2010,Poletto2013,Poletto2015}.
These factors, however, were found to be unimportant in other circumstances. For instance, the competition between two SIS processes starting at the same time on a static network was found to be unaffected by the network topology under the quenched mean field approximation \cite{Prakash2012} (where the spectral radius that rules the strength of spreading factorizes and disappears from the inequalities). An analytical formulation of the competition dynamics in the temporal case, based on the quenched mean field approximation would allow a direct comparison with the static case. The quenched mean field has been extended to temporal networks already, to compute the epidemic threshold \cite{Valdano2015}. The extension of this theory to the case of two competing pathogens is the subject of future research.

A second result is that strong fluctuations in the overall volume of contacts make the timing of emergence critical for the likelihood of replacement. Right after the peak of activity, the prevalence of the endemic strain is still sufficiently high to hamper the takeover by the emerging one. Chance of replacement is instead higher before the activity peak, when the prevalence of the endemic strain is at the minimum. Most importantly, the emergence time alters the shape of the competition diagram. When emergence occurs after the peak, any increase in transmissibility has limited effects on the probability of replacement, that is instead more affected by the infectious period. The interest on the competition diagrams stems from the fact that they can provide insights on possible mechanisms of selective pressure at the basis of pathogen evolution. Further consideration on this direction are interesting matter of research for the future.

The values of the average infectious period (around 6 hours for the endemic strain and between 5.5 hours and 50 hours for the emerging one) were chosen to be comparable with the time scale of network dynamics, that is dominated by daily activity cycles. This was motivated by the interest in probing the interacting dynamics between the two strain in a regime in which the network temporal structure is expected to have the highest degree of interference. In real cases, infectious durations are in general longer -- from around 1 day, as the case of common cold, up to several weeks for bacterial colonization and carriage in some cases. Exploring these regimes of the parameters has to be done in a future work, in order to understand the range of values where the main result found here still holds and the interference of the network dynamics is important.

Our work is just a preliminary attempt at uncovering how the complex interplay between network and pathogen time scales shape the rise and fixation of emerging strains. The theoretical interest in the problem is motivated by the importance of the emergence of new strains in disease ecology (pandemic strains, selection of multi-resistant bacteria) and other contexts (e.g. competition of meme and ideas). Future work is needed broaden the scope and applicability of our results.

\section*{Acknowledgements}
This work is the output of the Complexity72h workshop, held at IMT School in Lucca, 7-11 May 2018. \texttt{https://complexity72h.weebly.com/}.\\
S.C. acknowledges support from the DAIS-ITA project: "This research was sponsored by the U.S. Army Research Laboratory and the U.K. Ministry of Defence under Agreement Number W911NF-16-3-0001. The views and conclusions contained in this document are those of the authors and should not be interpreted as representing the official policies, either expressed or implied, of the U.S. Army Research Laboratory, the U.S. Government, the U.K. Ministry of Defence or the U.K. Government. The U.S. and U.K. Governments are authorized to reproduce and distribute reprints for Government purposes notwithstanding any copyright notation hereon."; M.G.L. acknowledges funding from the EU via H2020 Marie Sklodowska Curie project COSMOS, grant no. 642563; X.R.H. acknowledges support from the Ministerio de Educaci\'on, Cultura y Deporte of Spain, scholarship no. FPU16/05751; X.R.H. and E.O. acknowledge support from the Ministerio de Econom\'ia y Competitividad of Spain, project no. FIS2016-76830-C2-2- P (AEI/FEDER, UE); E.O. acknowledges support from a James S. McDonnell Foundation Scholar Award in Complex Systems; L.Z. acknowledges support from the Austrian Science Foundation, under FWF projects P29032; C.P. acknowledges funding from the program  Investissements d'Avenir, Sorbonne Universités.

\begin{figure}[htbp]
\centering
\includegraphics[width=0.8\textwidth]{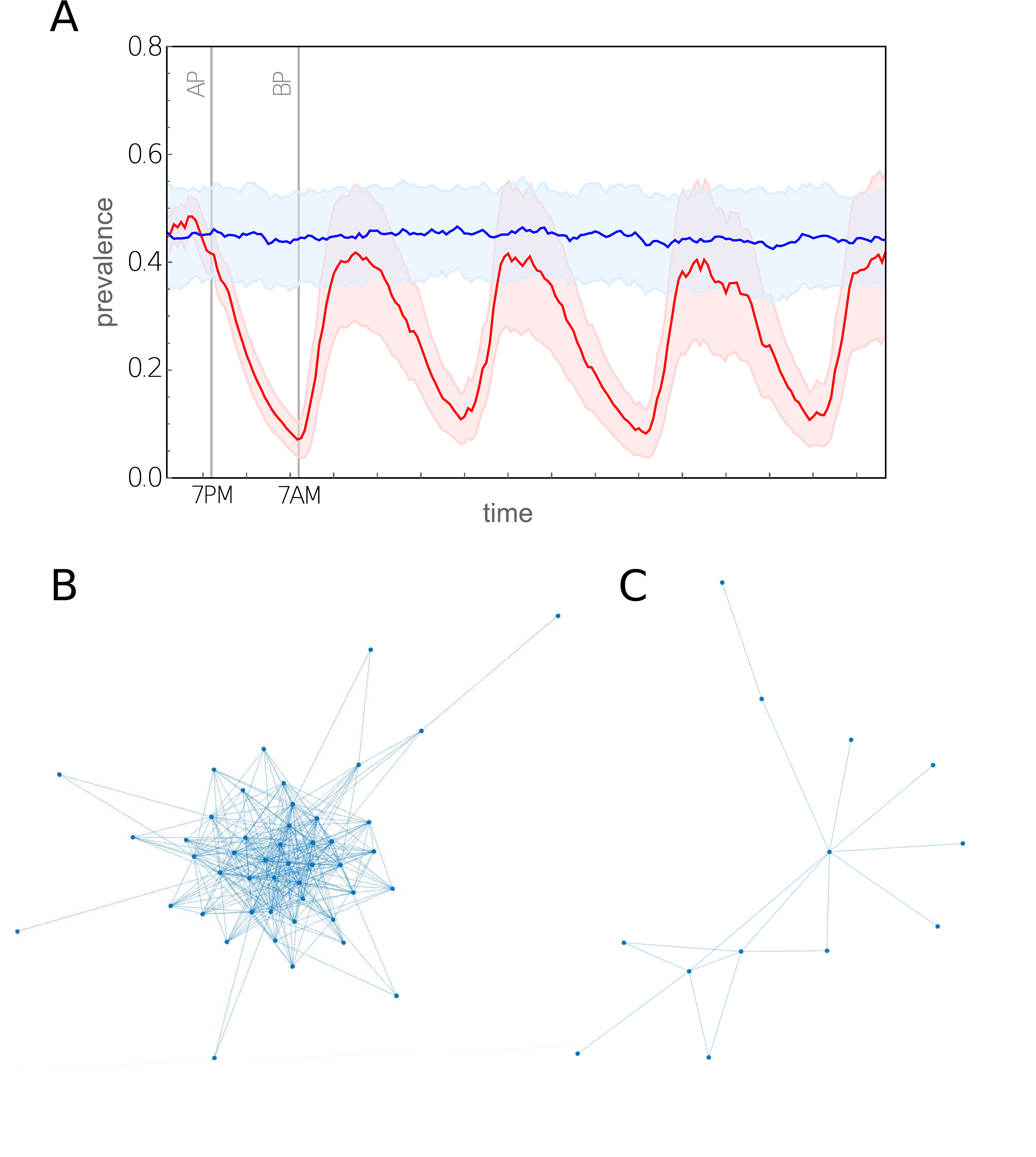}
\caption{\label{figure1}\textbf{Prevalence of the endemic strain in the temporal network and for the null model}. \textbf{a)} shows the average prevalence of wild-type strain (red solid line) and its standard deviation over 100 realizations (shaded red area), together with the average prevalence of emergent (blue solid line) and its standard deviation (blue shaded area). Two emergence times $t_{\mathrm{inj}}$ are denoted by vertical gray solid lines. The first time of emergence, at 7p.m., corresponds the emergence time after the peak $T_{AP}$, while 7a.m corresponds to  the emergence time before the peak $T_{BP}$. Panels \textbf{b)} and \textbf{c)} show each one, a sample of the contact network, at midday and midnight respectively.}
\end{figure}

\begin{figure}[htbp]
\centering
\includegraphics[width=\textwidth]{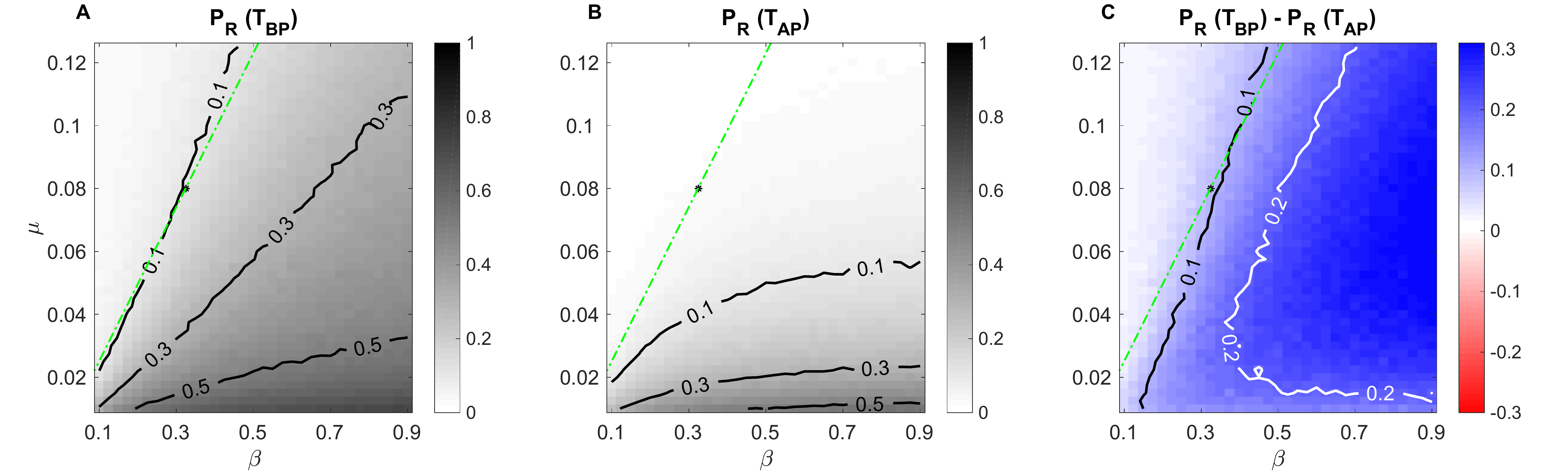}
\caption{\label{figure2}\textbf{Probability of replacement by emergent strain in the real temporal network, $P_{\mathrm{R}}$,  for several values of $(\beta,\mu)$  and different emergence times.} In \textbf{a)} the emergence time is $t_{\mathrm{inj}}=T_{\mathrm{BP}}.$ and in \textbf{b)} it corresponds to $t_{\mathrm{inj}}=T_{\mathrm{AP}}.$ \textbf{c)} shows the difference between probabilities of replacement at emergence times $t_{\mathrm{inj}}=T_{\mathrm{BP}}$ and $t_{\mathrm{inj}}=T_{\mathrm{AP}}$. The black asterisks indicate the values of the recovery probability ($\mu_0=0.08$) and the transmission probability ($\beta_0=0.325$) of the endemic  strain. The green lines show the values of $(\beta,\mu)$ where it is satisfied $\mu/\beta = \mu_0/\beta_0$. Contour lines are plotted in c) to for the sake of visualization.}
\end{figure}

\begin{figure}[htbp]
\centering
\includegraphics[width=1.0\textwidth]{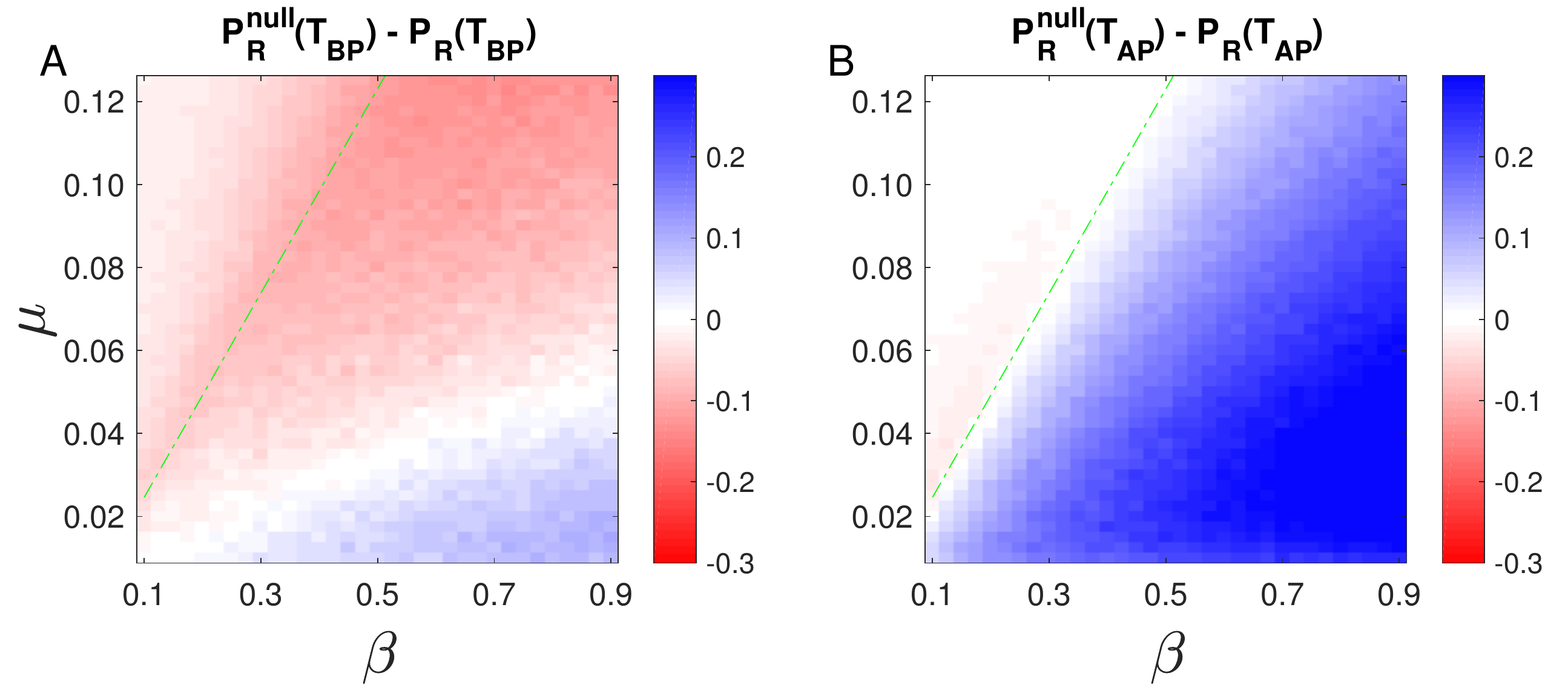}
\caption{\label{figure3} \textbf{Difference between probabilities of replacement, of the real temporal network and the null model for different emergence times.} \textbf{a)} shows the difference between the probabilities of replacement  of the null model  $P_{\mathrm{R}}^{\mathrm{null}}$, and the temporal network, $P_{\mathrm{R}}$, at emergence time $t_{\mathrm{inj}}=T_{\mathrm{BP} }$.  \textbf{b)} shows $(P_{\mathrm{R}}^{\mathrm{null}}-P_{\mathrm{R}})$ when emergence time is $t_{\mathrm{inj}}=T_{\mathrm{AP}}$. The highest relative replacement probability, 0.3, corresponds to the brightest blue color while the lowest value (-0.3) is indicated by the brightest red. The discontinuous green lines show the values of $(\beta,\mu)$ where it is satisfied $\mu/\beta = \mu_0/\beta_0$}
\end{figure}

\begin{figure}[htbp]
\centering
\includegraphics[width=0.6\textwidth]{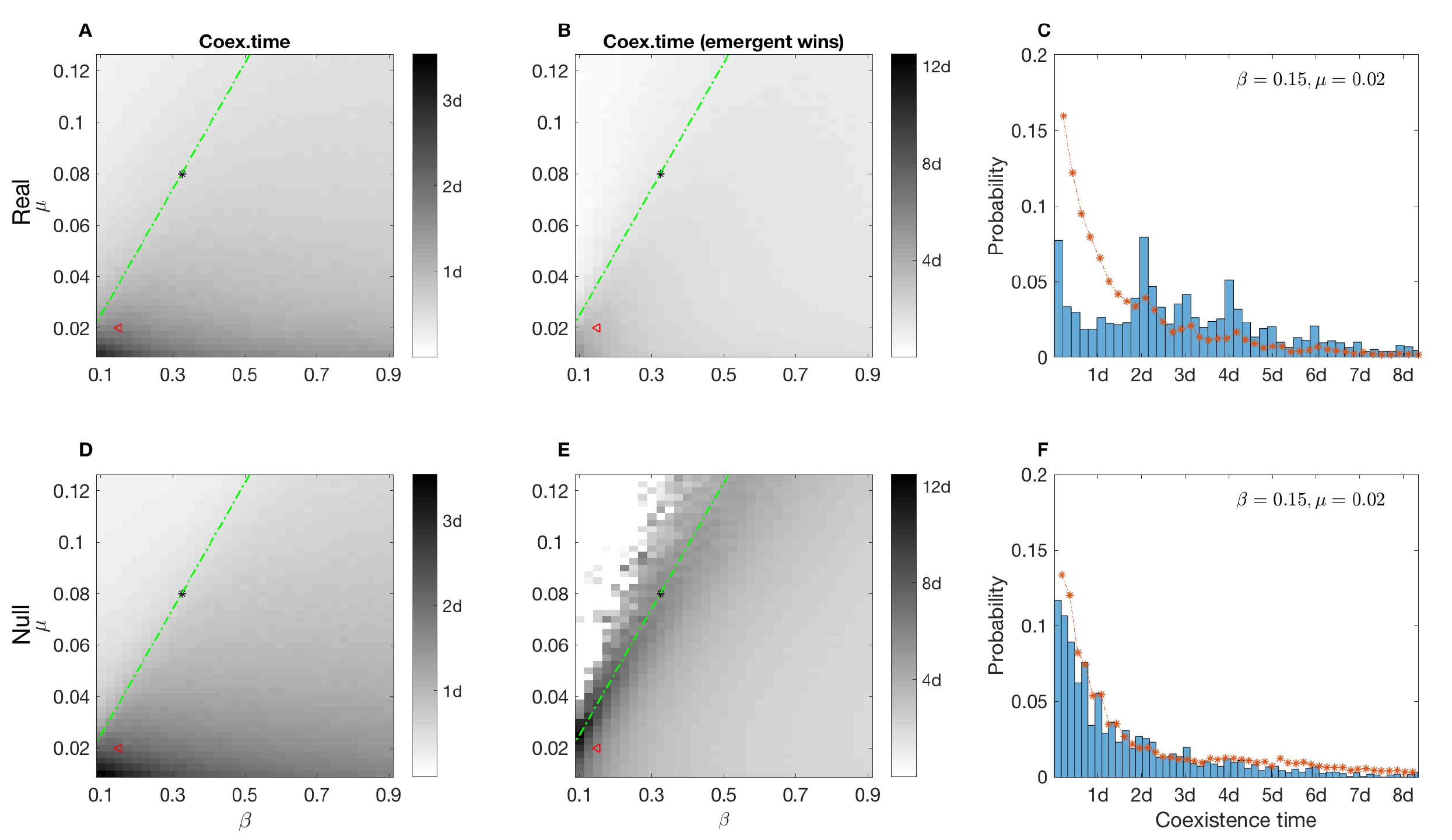}
\caption{\label{figure4}\textbf{Time of coexistence of wild-type and emergent strains in the real temporal network (top row) and in the null model (bottom row),} for several values of $(\beta,\mu)$ at emergence time $T_{BP}$.  \textbf{a)} and \textbf{d)} show the coexistence time independently of which strain eventually fixates, respectively, for the temporal network and the null model. \textbf{b)} and \textbf{e)} show the coexistence time given the condition that the emergent strain eventually replaces the wild-type, in the temporal network and in the null model respectively. The black asterisks indicate the values of the recovery probability ($\mu_0=0.08$) and the transmission probability ($\beta_0=0.325$) of the wild-type strain. The red triangles in the heat-maps correspond to the values of $(\beta=0.15,\mu=0.02)$ for which we show coexistence time distributions at the rightmost column. The bar plots in \textbf{c)} and \textbf{f)} show the distribution of coexistence times conditioned to the mutant strain overtaking the wild-type. Moreover, the distribution of coexistence times given that any of the two strains eventually prevails appears plotted in orange dotted line.}
\end{figure}

\end{document}